\documentclass[showpacs,preprintnumbers,amsmath,amssymb,prb]{revtex4}

\usepackage{graphicx}
\usepackage{dcolumn}
\usepackage{bm}

\begin{document}

\title{Elementary excitations in the ordered phase of spin-$\frac12$ $J_1$--$J_2$ model on square lattice}

\author{A.\ V.\ Syromyatnikov$^{1,2}$}
\email{asyromyatnikov@yandex.ru}
\author{A.\ Yu.\ Aktersky$^{1}$}
\email{aktersky@gmail.com}
\affiliation{$^1$National Research Center "Kurchatov Institute" B.P.\ Konstantinov Petersburg Nuclear Physics Institute, Gatchina 188300, Russia}
\affiliation{$^2$St.\ Petersburg State University, 7/9 Universitetskaya nab., St.\ Petersburg, 199034
Russia}

\date{\today}

\begin{abstract}

We use recently proposed four-spin bond-operator technique (BOT) to discuss spectral properties of frustrated spin-$\frac12$ $J_1$--$J_2$ Heisenberg antiferromagnet on square lattice at $J_2<0.4J_1$ (i.e., in the N\'eel ordered phase). This formalism is convenient for the consideration of low-lying excitations which appear in conventional approaches as multi-magnon bound states (e.g., the Higgs excitation) because separate bosons describe them in BOT. At $J_2=0$, the obtained magnon spectrum describes accurately available experimental data. However, calculated one-magnon spectral weights and the transverse dynamical structure factor (DSF) do not reproduce experimental findings quantitatively around the momentum ${\bf k}=(\pi,0)$. Then, we do not support the conjecture that the continuum of excitations observed experimentally and numerically near ${\bf k}=(\pi,0)$ is of the Higgs-magnon origin. Upon $J_2$ increasing, one-magnon spectral weights decrease and spectra of high-energy spin-0 and spin-1 excitations move down. One of spin-0 quasiparticles becomes long-lived and its spectrum merges with the magnon spectrum in the most part of the Brillouin zone at $J_2\approx0.3J_1$. We predict that the Higgs excitation and another spin-0 quasiparticle become long-lived around ${\bf k}=(\pi/2,\pi/2)$ at $J_2\agt0.3J_1$ and produce sharp anomalies in the longitudinal DSF.

\end{abstract}

\pacs{75.10.Jm, 75.10.-b, 75.10.Kt}

\maketitle

\section{Introduction}

Despite its simplicity and many experimental and theoretical efforts devoted to its investigation, spin-$\frac12$ Heisenberg antiferromagnet (HAF) on square lattice continues to attract much attention. This interest is stimulated greatly by the relevance of this model to physics of parent cuprate high temperature superconductors. \cite{monous} Of particular importance are magnetic excitations in spin-$\frac12$ HAF and their evolution in cuprates upon doping on the way from the antiferromagnetic (AF) insulating to the superconducting state. Spin excitations are considered now as one of the promising candidates to provide a "glue" for high temperature superconductivity. \cite{Tacon}

While properties of long-wavelength elementary excitations (magnons) in spin-$\frac12$ HAF on square lattice are well understood, \cite{chak,hydro,monous} the nature of short-wavelength magnons remains a subject of controversial debates. It is important to clarify this point in view of recent findings that short-wavelength spin excitations would play an important role in the spin-fluctuation-mediated pairing mechanism in high temperature superconductors. \cite{Tacon} It was observed both experimentally  in ${\rm Cu(DCOO)_2 \cdot 4D_2O}$ (CFTD) \cite{chris1,piazza} and numerically \cite{qmc2,sand,ser,spinon,piazza,pow1,pow2} that the magnon spectrum has a local minimum at ${\bf k}=(\pi,0)$ which is not reproduced quantitatively by analytical approaches including the spin-wave theory in the third order in $1/S$ (Refs.~\cite{igar2,syromyat}). Besides, a pronounced high-energy continuum of excitations arises in the transverse dynamical structure factor (DSF) at ${\bf k}=(\pi,0)$ having the form of a tail of the one-magnon peak. This high-energy tail was previously interpreted as an indication of an instability of the magnon having ${\bf k}=(\pi,0)$ with respect to a decay either on two spinons (Refs.~\cite{spinon,piazza,spinon2,spinon3}) or on another magnon and a Higgs excitation (Refs.~\cite{pow1,pow2,moess}). It is also proposed in the latter conjecture that a magnon attraction is in the origin of the local minimum in the spectrum. \cite{pow1,pow2,moess} 

Spin excitations in the so-called spin-$\frac12$ $J_1$--$J_2$ Heisenberg model are also of interest now. This model is an extension of the spin-$\frac12$ HAF on square lattice which contains along with the nearest-neighbor AF exchange coupling $J_1$ a frustrating next-nearest-neighbor exchange interaction $J_2$. Its Hamiltonian has the form
\begin{equation}
\label{ham}
{\cal H} = \sum_{\langle i,j \rangle}	{\bf S}_i{\bf S}_j 
+ 
J_2\sum_{\langle \langle i,j \rangle \rangle}	{\bf S}_i{\bf S}_j,
\end{equation}
where we put $J_1=1$. It was proposed that this model can describe doped cuprate superconductors in which a small concentration of holes appears in CuO planes. \cite{Inui} Some variants of the $J_1$--$J_2$ model have been used also to describe the weakened AF long-range order in iron-based high temperature superconductors. \cite{Dai2012} Besides, model \eqref{ham} has provided a convenient playground for the investigation of such novel types of many-body phenomena as quantum spin-liquid phases \cite{balents,vmc} and a novel universality class of phase transitions \cite{deconf,deconf2}.

It is generally believed that N\'eel ordered phases with AF vectors $(\pi,\pi)$ and $(\pi,0)$ (or $(0,\pi)$) arise at $J_2\alt0.4$ and $J_2\agt0.6$, respectively, and there is a magnetically disordered state in the intermediate region of $0.4\alt J_2\alt 0.6$. In spite of extensive investigations over the past three decades by various numerical and analytical methods, \cite{vmc,vmcjap,balents,fisher,darradi,richter,serplaq,serplaq2,serdim1,serdim2,serdim3,tensnet,tensnet2,ed,ed2,ed3,sw,sach1,sach2,mf,kotov,bond,bond2,wolfle} the nature of the nonmagnetic region remains unclear. While the disordered region is in the focus of attention now, the influence of the frustration on the N\'eel ordered phase at $J_2\alt0.4$ has not been discussed yet in every detail neither theoretically nor experimentally (although some suitable compounds with $J_2\approx0.2-0.3$ are known \cite{pbvo,vomo}).

We address in the present paper evolution of high-energy elementary excitations in the N\'eel state of $J_1$--$J_2$ HAF upon $J_2$ increasing at $0\le J_2<0.4$. We use a bond-operator technique (BOT) suggested by one of us \cite{ibot} which is suitable for describing both magnetically ordered and disordered states as well as transitions between them. This approach which is discussed briefly in Sec.~\ref{method} is guided by the idea to increase the unit cell in order to take into account all spin degrees of freedom in it. There are extra bosons in the bosonic representation of spin operators in the unit cell which describe elementary excitations arising in conventional approaches as bound states of ordinary quasiparticles (magnons or triplons). In particular, in BOT with four spins in the unit cell which was suggested for the ordered phase in spin-$\frac12$ HAF, there are separate bosons describing the amplitude (Higgs) excitation and a spin-0 quasiparticle named singlon. \cite{ibot} The latter is responsible for the anomaly in Raman intensity in the $B_{1g}$ symmetry observed, e.g., in layered cuprates. \cite{ibot} The proposed variant of BOT allows a regular expansion of physical observables in powers of $1/n$, where $n$ is the maximum number of bosons which can occupy a unit cell (physical results correspond to $n=1$). The spin commutation algebra is fulfilled for any $n>0$. By comparison with other available numerical and experimental results, it was demonstrated \cite{ibot} that first $1/n$ corrections make the main renormalization to the staggered magnetization, the ground-state energy, and energies of quasiparticles. On the other hand, quasiparticles damping appears to be too rough in the first order in $1/n$ as it is the order in which first nonzero corrections to the damping appear.

As it was obtained in our previous paper \cite{ibot}, the spectrum of magnons is reproduced quite accurately at $J_2=0$ within the first order in $1/n$ even around ${\bf k}=(\pi,0)$. We demonstrate in Sec.~\ref{specsec} that one-magnon spectral weights are in a very good agreement in the whole Brillouin zone (BZ) with the experiment in CFTD \cite{chris1,piazza} except for the neighborhood of ${\bf k}=(\pi,0)$. Consideration within our approach could support the Higgs-magnon origin of the continuum of excitations above the magnon peak at ${\bf k}=(\pi,0)$ because the magnon and the amplitude excitations appear explicitly in BOT. However, we observe a very weak Higgs-magnon continuum in the first order in $1/n$. Besides, we find that calculated one-magnon spectral weights are overestimated near ${\bf k}=(\pi,0)$. Thus, we do not confirm the Higgs-magnon origin of the continuum.

We examine the effect of the frustration in Sec.~\ref{j2ordered} and demonstrate that magnon spectral weights are reduced upon $J_2$ rising. The deviation around ${\bf k}=(\pi,0)$ becomes more pronounced of the calculated magnon spectrum from that found in the second order in $1/S$. We observe that spectra of all quasiparticles move down when $J_2$ increases. However spectra of high-energy spin-0 and spin-1 elementary excitations move faster than the magnon spectrum. As a result, the singlon spectrum merges with the magnon one in the most part of BZ at $J_2\approx0.3$. Besides, the Higgs excitation, another spin-0 quasiparticle and a spin-1 elementary excitation become also very close to the magnon spectrum at ${\bf k}=(\pi/2,\pi/2)$ and $J_2\agt0.3$. Then, we predict that the Higgs and the spin-0 excitations produce distinct anomalies around ${\bf k}=(\pi/2,\pi/2)$ in the longitudinal DSF whose spectral weights are also calculated. The spin-1 quasiparticle produces an anomaly in the transverse DSF near ${\bf k}=(\pi/2,\pi/2)$ whose spectral weight is more than an order of magnitude smaller than the spectral weight of the one-magnon peak.

We provide a summary and a conclusion in Sec.~\ref{conc}. An appendix is also added with details of calculations.

\section{Model and technique}
\label{method}

We double the unit cell in two directions so that there are four spins in the unit cell and introduce 15 Bose-operators in the BOT formulated by one of us in Ref.~\cite{ibot}. Two bosons describe high-energy spin-2 excitations, eight Bose-operators stand for spin-1 quasiparticles four of which are magnons, and there are five spin-0 excitations two of which are two parts of the amplitude mode and one boson describes the singlon.

\begin{figure}
\includegraphics[scale=0.4]{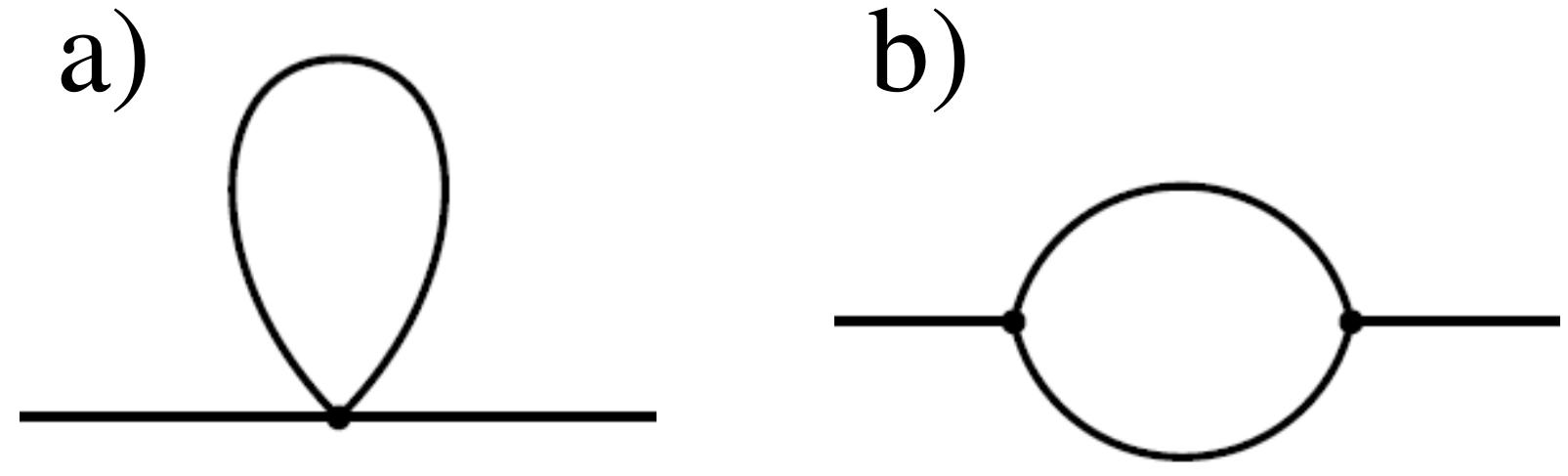}
\caption{Diagrams giving corrections of the first-order in $1/n$ to self-energy parts.
\label{diag}}
\end{figure}

\begin{figure}
\includegraphics[scale=0.1]{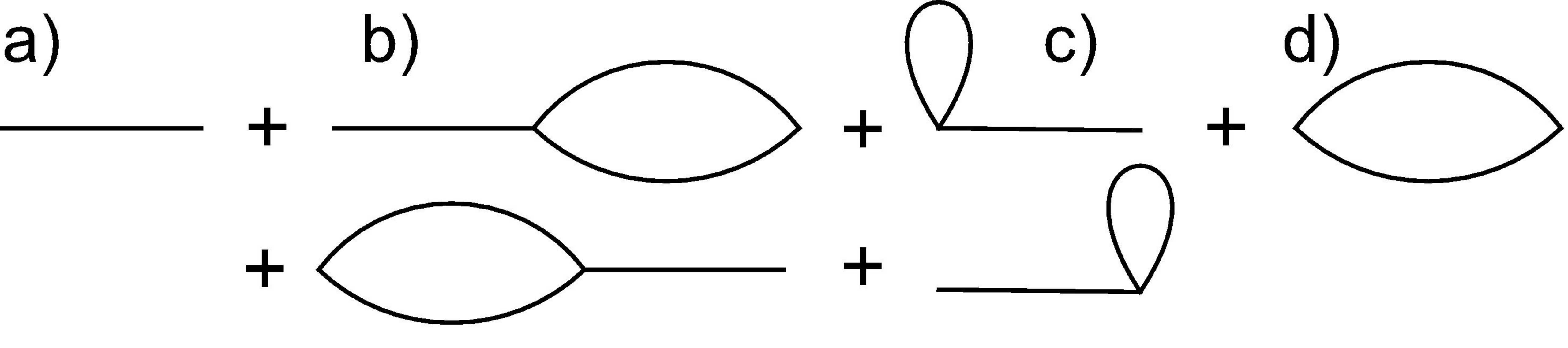}
\caption{Diagrams for spin susceptibilities \eqref{dsf} to be taken into account in the first order in $1/n$.
\label{chifig}}
\end{figure}

We calculate spin susceptibilities (SSs)
\begin{equation}
\label{dsf}
\chi_{\alpha\beta}(\omega,{\bf k}) = 
i\int_0^\infty dt 
e^{i\omega t}	
\left\langle \left[ S^\alpha_{\bf k}(t), S^\beta_{-\bf k}(0) \right] \right\rangle,
\end{equation}
where spin operators read in our terms as 
$
S^\gamma_{\bf k} = S^\gamma_{1\bf k} + S^\gamma_{2\bf k}e^{-ik_y/2} + S^\gamma_{3\bf k}e^{-i(k_x+k_y)/2} + S^\gamma_{4\bf k}e^{-ik_x/2}
$, the double distance between nearest neighbor spins is set to be equal to unity here and spins in the unit cell are enumerated clockwise starting from its left lower corner. We use the representation of spins components $S^\gamma_{1,2,3,4}$ via Bose-operators proposed in Ref.~\cite{ibot} and calculate SSs within the first order in $1/n$ by the conventional diagram technique as it is explained in Ref.~\cite{ibot}. In particular, we calculate diagrams shown in Figs.~\ref{diag} and \ref{chifig} to find self-energy parts and SSs, respectively.

\begin{figure}
\includegraphics[scale=0.3]{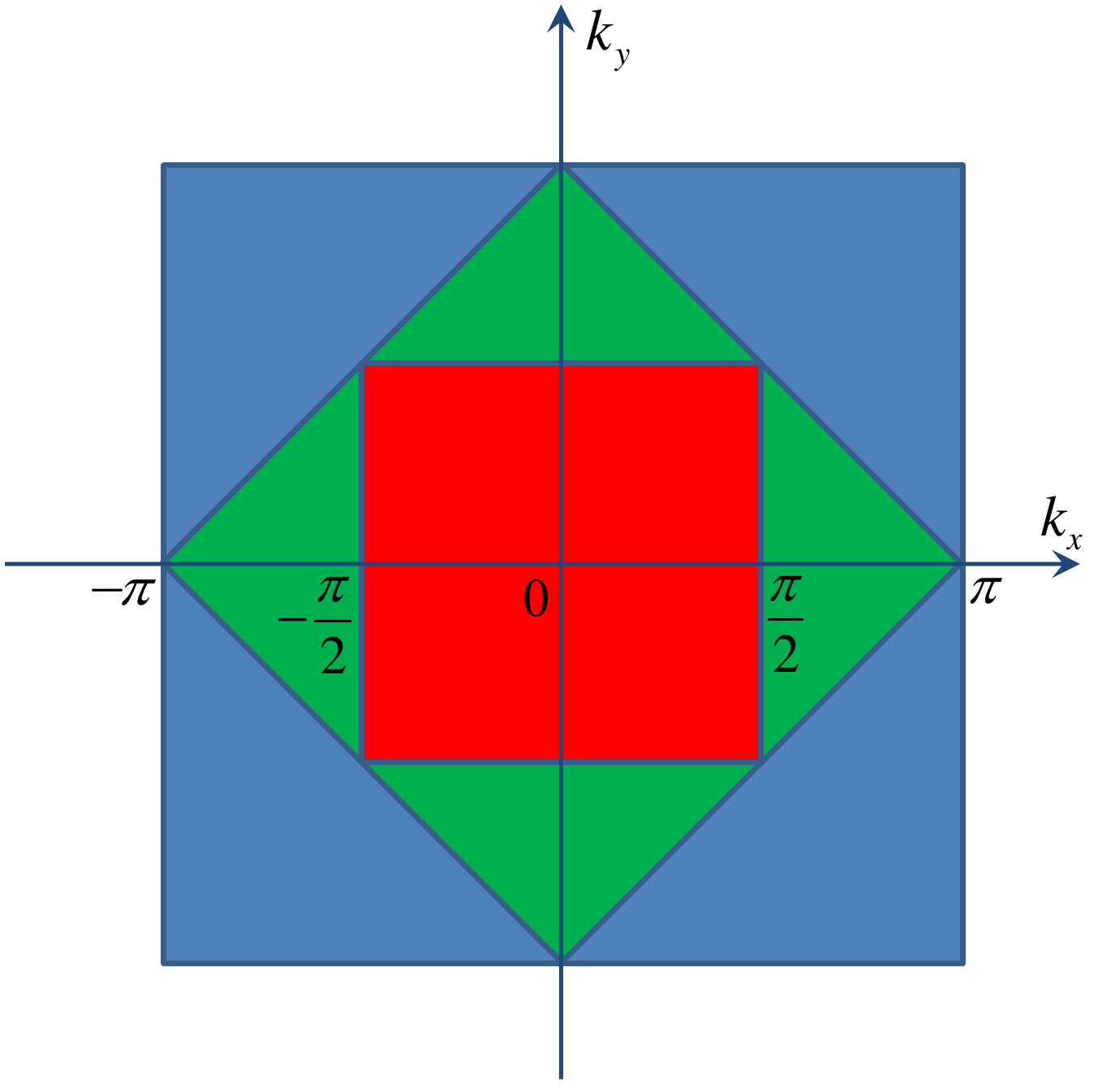}
\caption{The chemical and magnetic Brillouin zones (BZs) are presented (the largest and the middle squares, respectively) for the simple square lattice. The distance between nearest lattice sites is set to be equal to unity. The smallest (red) square and the green area are the first and the second BZs, correspondingly, in the case of four sites in the unit cell having the form of a plaquette.
\label{bz}}
\end{figure}

Transverse $\chi_{+-}(\omega,{\bf k})$ and longitudinal $\chi_{zz}(\omega,{\bf k})$ SSs are expressed in the leading (zeroth) order in $1/n$ via Green's functions of Bose-operators describing spin-1 and spin-0 excitations, respectively. In the first order in $1/n$, 
denominators ${\cal D}(\omega,{\bf k})$ of SSs can be represented as an expansion of the following expression up to terms of ${\cal O}(1/n)$ (see also Appendix~\ref{app}):
\begin{eqnarray}
\label{denom}
{\cal D}(\omega,{\bf k}) &=& 
\left( 1 + \frac1n\delta_0(\omega,{\bf k}) \right)	
\left(\omega^2 - \left(\epsilon_{1\bf k}^{(0)} + \frac1n\delta_1(\omega,{\bf k})\right)^2 \right)	
\left(\omega^2 - \left(\epsilon_{2\bf k}^{(0)} + \frac1n\delta_2(\omega,{\bf k})\right)^2 \right)\nonumber\\	
&&\times\left(\omega^2 - \left(\epsilon_{3\bf k}^{(0)} + \frac1n\delta_3(\omega,{\bf k})\right)^2 \right)	
\left(\omega^2 - \left(\epsilon_{4\bf k}^{(0)} + \frac1n\delta_4(\omega,{\bf k})\right)^2 \right),
\end{eqnarray}
where $\epsilon_{1,2,3,4\bf k}^{(0)}$ are bare spectra of elementary excitations ($\epsilon_{1\bf 0}^{(0)}<\epsilon_{2\bf 0}^{(0)}<\epsilon_{3\bf 0}^{(0)}<\epsilon_{4\bf 0}^{(0)}$) and $\delta_{0,1,2,3,4}(\omega,{\bf k})$ are functions composed of self-energy parts which we do not present here due to their cumbersomeness.
\footnote{
Notice that corrections of the first order in $1/n$ to parameters $\alpha$ and $\beta$ introduced in Ref.~\cite{ibot} also contribute to $\delta_{1,2,3,4}(\omega,{\bf k})$ as well as to the numerator of the diagram shown in Fig.~\ref{chifig}(a).} 
The diagram shown in Fig.~\ref{chifig}(d) contributes to a background of SSs. 
Evidently, $\epsilon_{i\bf k}=\epsilon_{i\bf k}^{(0)}+\delta_i(\omega=\epsilon_{i\bf k}^{(0)})$, where $i=1,2,3,4$, give renormalized spectra in the first order in $1/n$. It is demonstrated in Ref.~\cite{ibot} that the residue of bare SSs at $\omega=\epsilon_{1\bf k}^{(0)}$ is strongly suppressed in the green area shown in Fig.~\ref{bz} whereas the residue at $\omega=\epsilon_{2\bf k}^{(0)}$ is strongly suppressed in the red area. Thus, $\epsilon_{1\bf k}$ and $\epsilon_{2\bf k}$ are two parts of quasiparticle spectra which meet at the border of the green and the red areas in the magnetic BZ (see Fig.~\ref{bz}). Then, $\epsilon_{1\bf k}$ and $\epsilon_{2\bf k}$ are two parts of spectra of magnons and the amplitude mode in the case of $\chi_{+-}(\omega,{\bf k})$ and $\chi_{zz}(\omega,{\bf k})$, respectively. In $\chi_{+-}(\omega,{\bf k})$, $\epsilon_{3\bf k}$ and $\epsilon_{4\bf k}$ are bare spectra of high-energy spin-1 excitations which can arise, e.g., in the conventional spin-wave formalism as bound states of three magnons (i.e., as poles of a three-particle vertex). In $\chi_{zz}(\omega,{\bf k})$, $\epsilon_{3\bf k}$ and $\epsilon_{4\bf k}$ are spectra of high-energy spin-0 excitations which correspond to two-magnon bound states in common approaches.

In addition to spin-0 excitations whose spectra are determined by poles of $\chi_{zz}(\omega,{\bf k})$, there is a special spin-0 mode which is purely singlet in phases with singlet ground states. \cite{ibot} For short, we call it singlon in Ref.~\cite{ibot} bearing in mind, however, that it is not singlet in the ordered phase. We demonstrated in Ref.~\cite{ibot} that singlon spectrum arises as a pole in bond-bond correlators. In particular, the Raman intensity in the $B_{1g}$ symmetry is expressed via the singlon Green's function at $\bf k=0$ which describes the so-called two-magnon asymmetric peak observed, e.g., in layered cuprates. \cite{ibot} Within the first order in $1/n$, the position of the peak is accurately reproduced, whereas the peak width is underestimated by roughly a factor of 3.

We find below one-magnon spectral weights by calculating the transverse DSF
\begin{equation}
\label{sperp}
{\cal S}_\perp(\omega,{\bf k}) = \frac{1}{2\pi}{\rm Im}
\left(\chi_{+-}(\omega,{\bf k}) + \chi_{-+}(\omega,{\bf k})\right)
= \frac{1}{\pi}{\rm Im}
\left(\chi_{xx}(\omega,{\bf k}) + \chi_{yy}(\omega,{\bf k})\right).
\end{equation}
Spectral weights of spin-0 quasiparticles are found from the longitudinal DSF ${\cal S}_\|(\omega,{\bf k}) = \frac{1}{\pi}{\rm Im} \chi_{zz}(\omega,{\bf k})$.

\section{Spectral properties at $J_2=0$}
\label{specsec}

\begin{figure}
\includegraphics[scale=0.8]{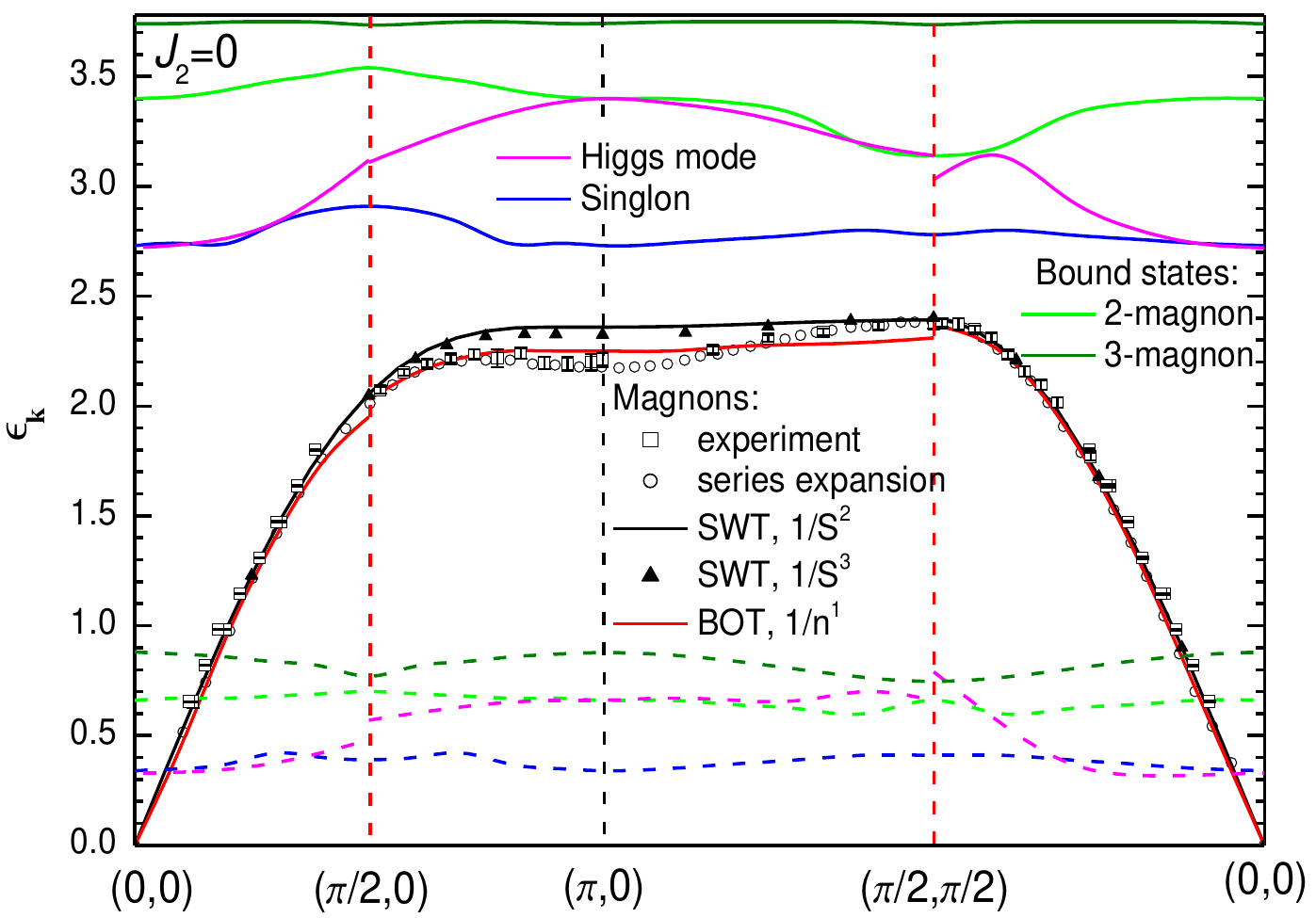}
\caption{Spectra of low-energy elementary excitations in the spin-$\frac12$ HAF on square lattice found using the four-spin bond-operator technique (BOT) in the first order in $1/n$. Also shown are magnon spectra obtained by series expansion around the Ising limit \cite{ser}, within the spin-wave theory (SWT) in the second \cite{igar,igar2} and in the third \cite{syromyat} orders in $1/S$, and neutron scattering experiment in CFTD \cite{chris1,piazza}. Borders of the first BZ with four spins in the unit cell are shown by red vertical lines (see Fig.~\ref{bz}). Small jumps in the magnon and in the Higgs mode spectra on the red vertical lines are an artifact of the first order in $1/n$ as it is explained in the text. Dashed lines correspond to the damping of quasiparticles whose energies are drawn by solid lines of the same color. Modes denoted as "bound states" are described in BOT by separate bosons while they would appear, e.g., in the spin wave theory as two- and three-magnon bound states.
\label{spec_ss}}
\end{figure}

Spectra of low-energy elementary excitations found within the first order in $1/n$ at $J_2=0$ are shown in Fig.~\ref{spec_ss}. It is seen that the singlon and the Higgs excitations are moderately damped in the first order in $1/n$. Besides, singlons lie below the amplitude mode in the major part of BZ. Notice that the spectrum of magnons is in a good quantitative agreement with previous numerical and experimental results in the whole BZ except for the neighborhood of borders between the red and the green areas shown in Fig.~\ref{bz}. Besides, there are small jumps in the magnon and in the Higgs mode spectra on the borders between the red and the green areas which should vanish after taking into account $1/n$ corrections of further orders. \cite{ibot} Remarkably, the observed magnon spectrum is in the quantitative agreement with experimental results even near ${\bf k}=(\pi,0)$ (see Fig.~\ref{spec_ss}): $\epsilon_{2\bf k}=2.23+0.02/n$ at this momentum. 

\begin{figure}
\includegraphics[scale=0.7]{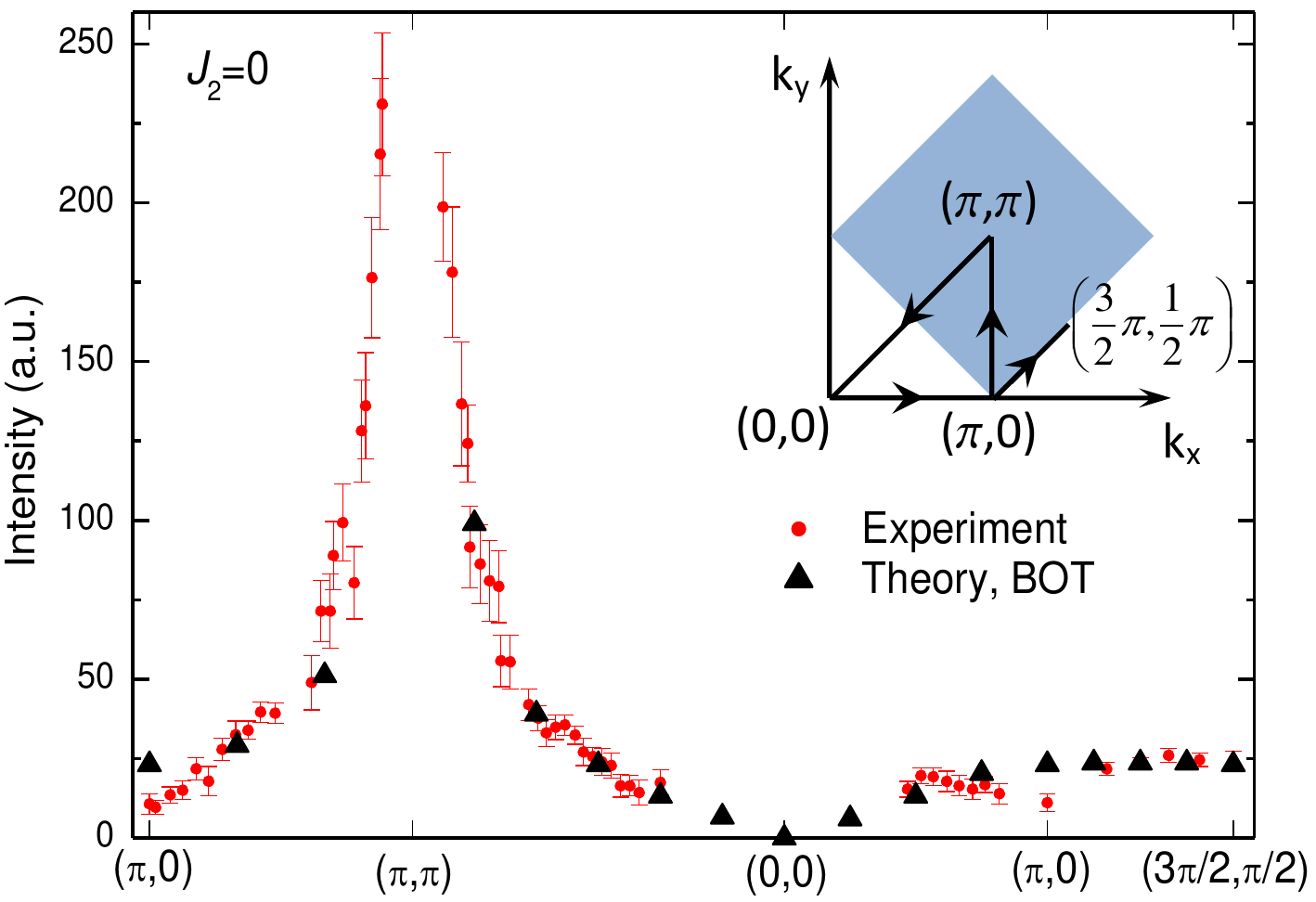}
\caption{One-magnon spectral weights obtained experimentally in CFTD \cite{chris1} and by the bond-operator technique (BOT) in the first order in $1/n$ (present study). Theoretical results are found using Eq.~\eqref{sperp} and they are multiplied by a common factor to fit the experimental data. 
\label{weights}}
\end{figure}

The spectral weight of the magnon pole at ${\bf k}=(\pi,0)$ found in the first order in $1/n$ by taking into account diagrams shown in Figs.~\ref{diag} and \ref{chifig}(a)--(c) reads as 
\begin{equation}
\label{weightpi0}
{\cal W}_m\Bigl(J_2=0,{\bf k}=(\pi,0)\Bigr) =	0.44 - \frac1n 0.02.
\end{equation}
Eq.~\eqref{weightpi0} gives $0.42$ at $n=1$ which should be compared with $0.43$, $0.31$, and $0.28$ found using the continuous similarity transformation (CST) technique (Ref.~\cite{pow2}), in the second order in $1/S$ (Ref.~\cite{igar2}), and by the series expansion (Ref.~\cite{ser}), respectively.
The spectral weight of the magnon pole at ${\bf k}=(\pi/2,\pi/2)$ reads as
\begin{equation}
\label{weightpipi}
{\cal W}_m\left(J_2=0,{\bf k}=\left(\frac\pi2,\frac\pi2\right)\right) =	0.44 - \frac1n 0.02.
\end{equation}
One gets $0.42$ from Eq.~\eqref{weightpipi} at $n=1$ 
which should be compared with $0.58$, $0.31$, and $0.35$ found using CST (Ref.~\cite{pow2}), in the second order in $1/S$ (Ref.~\cite{igar2}), and by the series expansion (Ref.~\cite{ser}), respectively.
It is seen from Fig.~\ref{spec_ss} that two magnon modes (describing by different bosons) exist at ${\bf k}=(\pi/2,\pi/2)$ which should merge upon taking into account corrections of all orders in $1/n$. Each of these modes produces a peak in the transverse DSF at ${\bf k}=(\pi/2,\pi/2)$ in the first order in $1/n$ so that the total spectral weight of these peaks stands in Eq.~\eqref{weightpipi}. One-magnon spectral weights in other representative points of BZ are shown in Fig.~\ref{weights}. A good quantitative agreement is seen between the experiment in CFTD \cite{chris1} and BOT in the whole BZ except for the vicinity of ${\bf k}=(\pi,0)$.

To discuss properties of magnons around ${\bf k}=(\pi,0)$, we calculate $\chi_{+-}(\omega,{\bf k})$ and $\chi_{-+}(\omega,{\bf k})$ in the first order in $1/n$. Contributions to these SSs from diagrams shown in Figs.~\ref{chifig}(a)--(c) contain denominator ${\cal D}(\omega,{\bf k})$ which can be represented in the first order in $1/n$ as a result of expansion of Eq.~\eqref{denom} up to terms of the first order in $1/n$, where $\epsilon_{1\bf 0}^{(0)}=0$ and $\epsilon_{2\bf 0}^{(0)}=2.23$ are bare energies of two magnons (at $\bf k=0$ and at ${\bf k}=(\pi,0)$, respectively), $\epsilon_{3\bf 0}^{(0)}=3.98$ and $\epsilon_{4\bf 0}^{(0)}=4.42$ are bare energies of spin-1 excitations at ${\bf k=0}$ (three-magnon bound states). We observe that self-energy parts acquire imaginary parts at $\omega$ greater than 2.43 (the bare energy of the Higgs excitation with $\bf k=0$) which originate from the diagram presented in Fig.~\ref{diag}(b). A detailed analysis shows that the major contribution to the imaginary parts arises due to the decay on a long-wavelength Higgs excitation and a long-wavelength magnon. However, the imaginary parts are very small at $2.43<\omega<3$ of self-energy parts corresponding to magnons with the spectrum $\epsilon_{2\bf k}$. In particular, ${\rm Im}\delta_{0,1,3,4}(\omega,{\bf 0})$ are pronounced at $2.43<\omega<3$ whereas ${\rm Im}\delta_2(\omega,{\bf 0})$ is negligible (see Eq.~\eqref{denom}). Accurate expansion of $\chi_{+-}(\omega,{\bf k}) + \chi_{-+}(\omega,{\bf k})$ up to terms of the first order in $1/n$ shows that self-energy parts with large imaginary parts from denominator cancel those from numerator. Imaginary parts of loops in diagrams presented in Figs.~\ref{chifig}(b) and \ref{chifig}(d) are also negligible at $\omega<3$. Thus, our results do not support the conjecture that the continuum in ${\cal S}_\perp(\omega,{\bf k})$ at ${\bf k}=(\pi,0)$ is of the magnon-Higgs type. 

\section{N\'eel phase at $0<J_2<0.4$}
\label{j2ordered}

\begin{figure}
\includegraphics[scale=0.75]{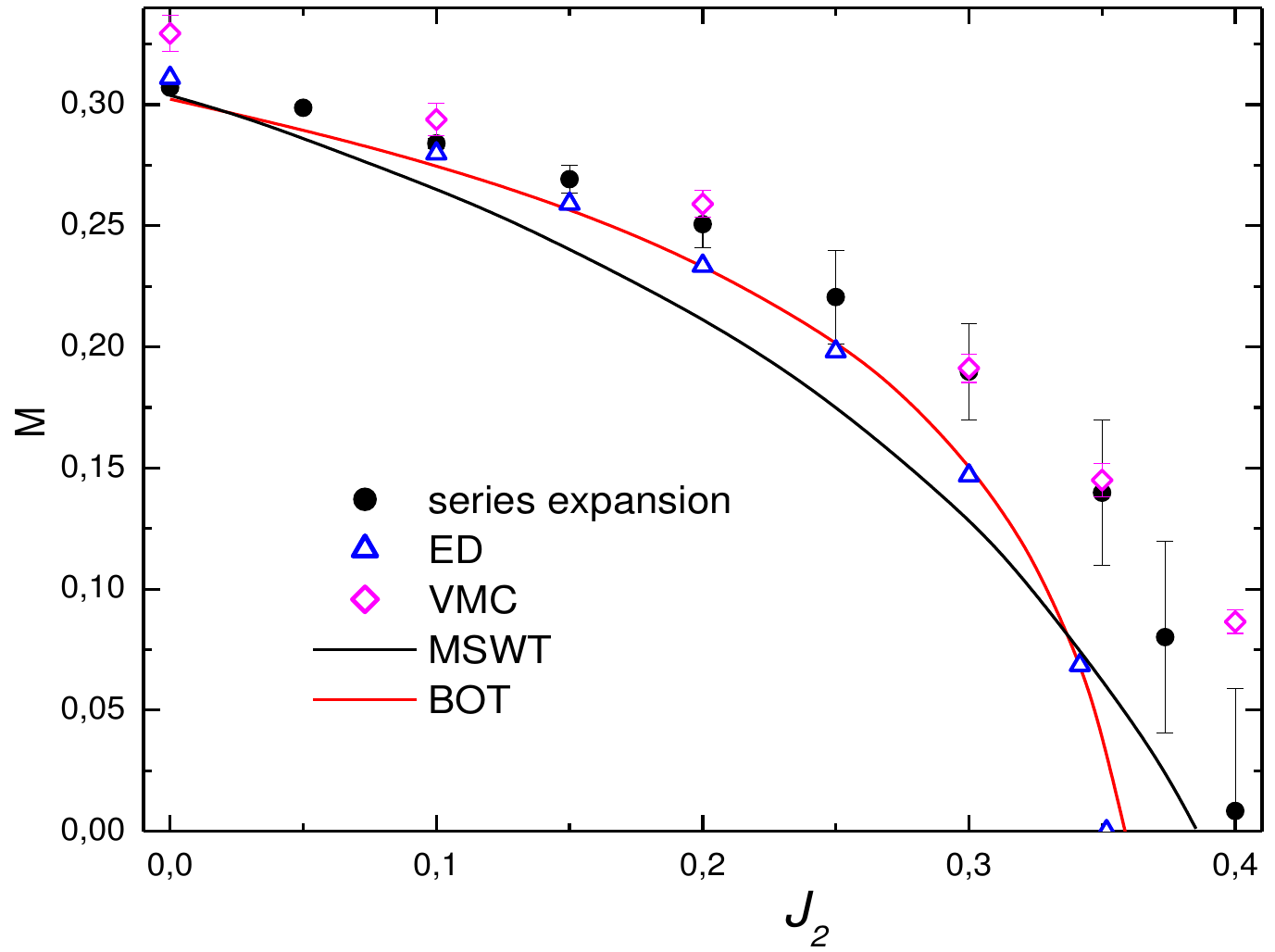}
\caption{Staggered magnetization $M$ found using the series expansion around the Ising limit \cite{serj1j2}, the exact diagonalization of finite clusters with extrapolation to the thermodynamic limit (ED) \cite{ed}, variational Monte Carlo simulations (VMC) \cite{vmcjap}, a modified spin-wave theory (MSWT) \cite{mswt}, and BOT in the first order in $1/n$ (present study).
\label{mfig}}
\end{figure}

We discuss in this section the N\'eel phase in the $J_1$--$J_2$ HAF using the BOT within the first order in $1/n$. The staggered magnetization $M$ shown in Fig.~\ref{mfig} is found as it is done in Ref.~\cite{ibot} for $J_2=0$. It is seen that our results are in good agreement with some other numerical findings at $J_2<0.3$. In particular, one obtains $M=0.301$ at $J_2=0$ in the first order in $1/n$ which is very close to the value of $\approx0.3$ observed before by many methods \cite{monous}. One obtains for the critical value of $J_2$ at which the order parameter vanishes
\begin{equation}
J_{2c} = 0.42 -\frac1n 0.06
\end{equation}
that gives $J_{2c} = 0.36$ at $n=1$ in agreement with many previous results showing $J_{2c}\approx0.4$.

\begin{figure}
\includegraphics[scale=0.75]{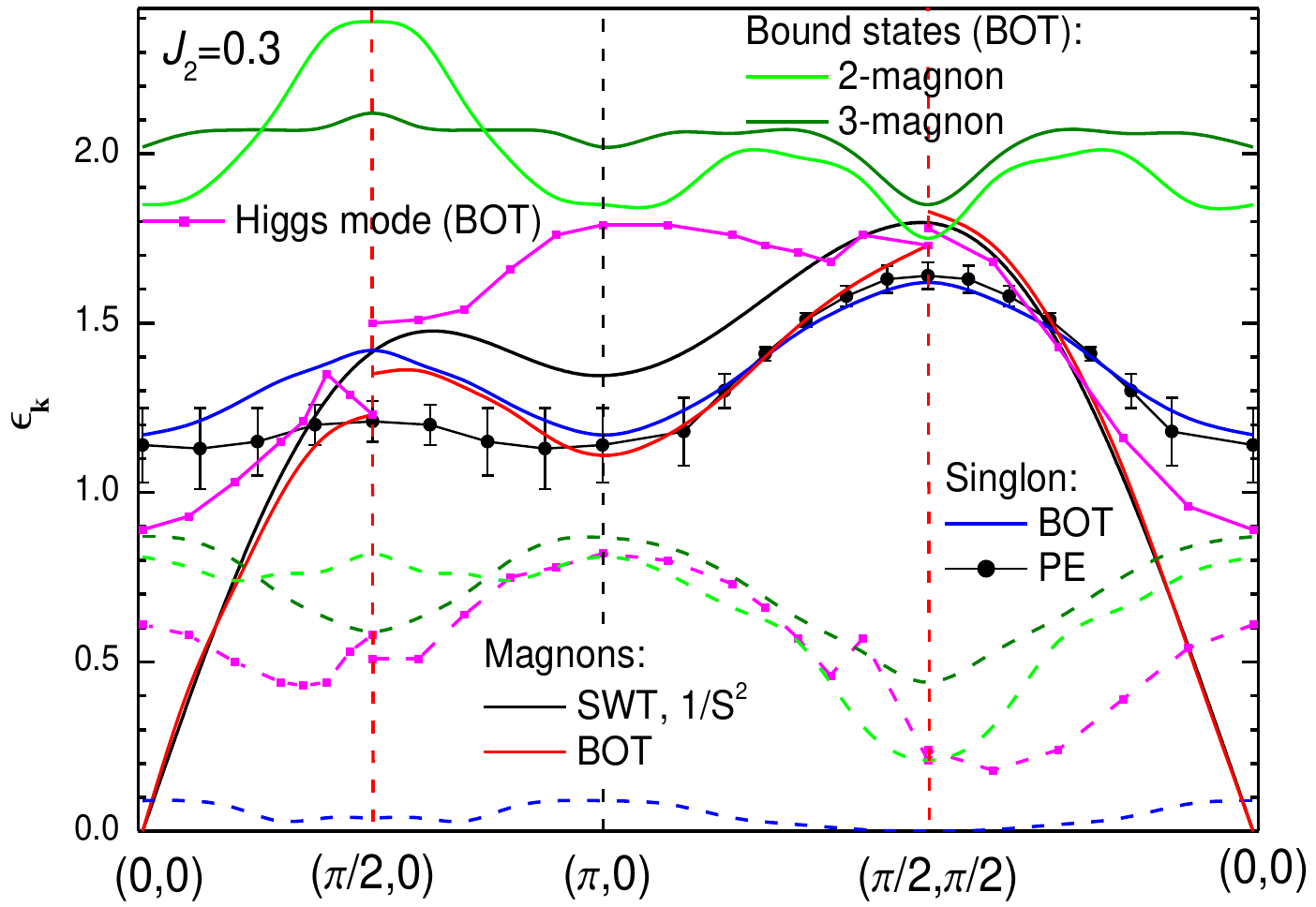}
\caption{Same as in Fig.~\ref{spec_ss} but for the $J_1$--$J_2$ model \eqref{ham} on square lattice at $J_2=0.3$. Also shown is the spectrum of singlons found using the plaquette expansion (PE) by the method proposed in Refs.~\cite{singlon,singlon2}.
\label{spec_j1j203}}
\end{figure}

Obtained spectra of low-lying elementary excitations are shown in Fig.~\ref{spec_j1j203} for $J_2=0.3$ (cf.\ Fig.~\ref{spec_ss}). Fig.~\ref{spec_j1j203} illustrates our observation that the deviation of the magnon spectrum found using BOT from that obtained in the second order in $1/S$ becomes more pronounced near ${\bf k}=(\pi,0)$ upon $J_2$ increasing. Notice that second-order $1/S$-corrections give a negligibly small renormalization of the magnon spectrum at all $J_2$. \cite{igar2,syromyat,kin}

Upon $J_2$ increasing, spectra of high-energy elementary excitations move down faster than energies of magnons. In particular, Fig.~\ref{spec_j1j203} shows that the spectrum of singlons, who remain the lower spin-0 excitations in the main part of BZ, merges with the spectrum of high-energy magnons at $J_2\approx0.3$. Besides, the singlon damping (which appears mainly due to singlon decay into two spin-1 excitations) decreases fast as $J_2$ rises so that singlons turn out to be long-lived quasiparticles at $J_2\approx0.3$, as it is seen from Fig.~\ref{spec_j1j203}. Spikes in the Higgs mode damping accompanied by abrupt changes in its energy is the appearance of the Van Hove singularities from the two-magnon density of states (similar anomalies were observed, e.g., in magnon spectra in the first order in $1/S$ in non-collinear magnets \cite{zhito,chub_triang} and in the Higgs mode spectrum in the Heisenberg bilayer model \cite{ibot}).

\begin{figure}
\includegraphics[scale=0.75]{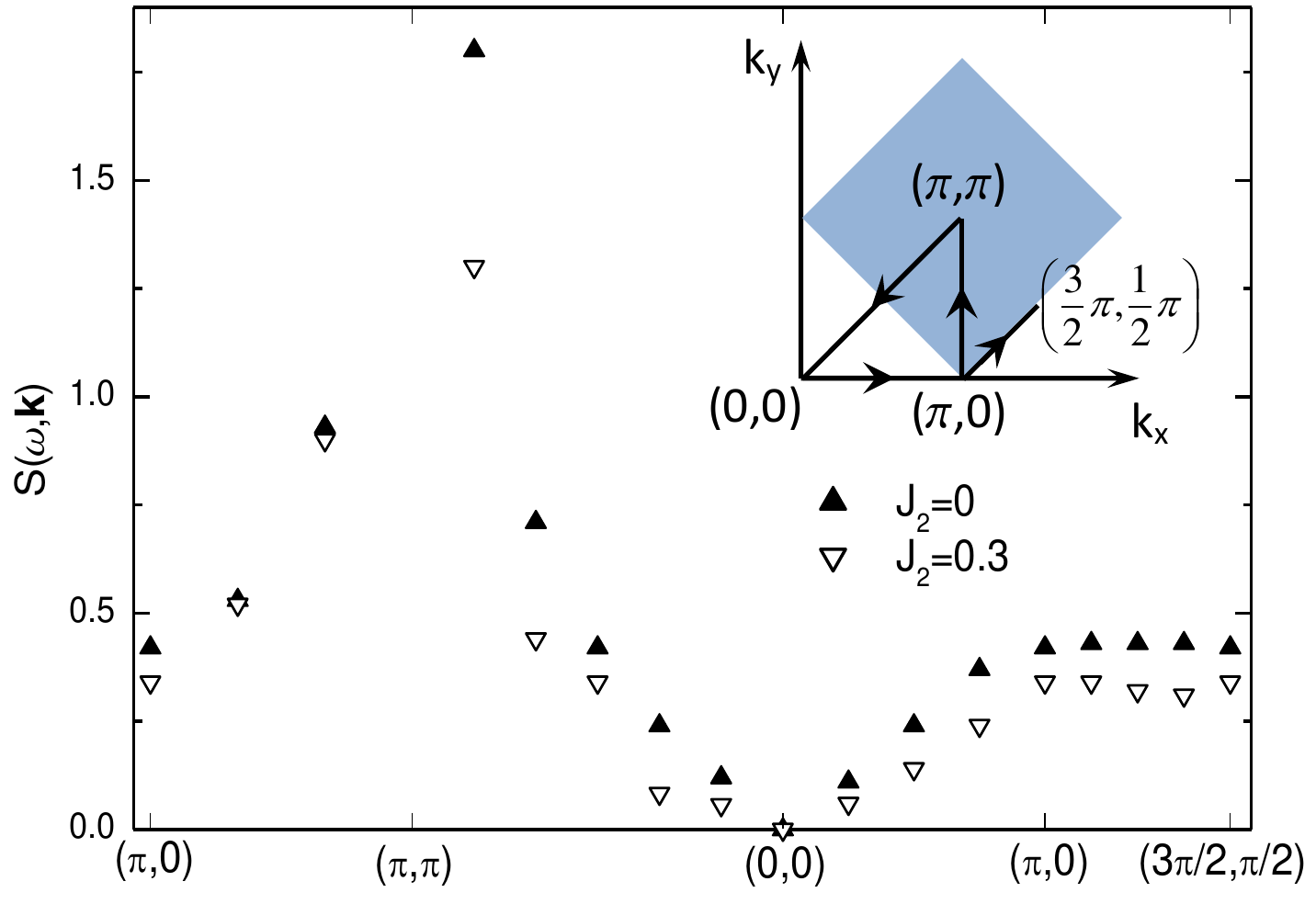}
\caption{One-magnon spectral weights obtained using Eq.~\eqref{sperp} in the first order in $1/n$ for $J_2=0$ and $J_2=0.3$.
\label{weights03fig}}
\end{figure}

As it is seen from Fig.~\ref{weights03fig}, one-magnon spectral weights decrease upon $J_2$ increasing. In particular, their values at ${\bf k}=(\pi,0)$ and ${\bf k}=(\pi/2,\pi/2)$ have the form (cf. Eqs.~\eqref{weightpi0} and \eqref{weightpipi})
\begin{eqnarray}
\label{weightpi0m}
{\cal W}_m \left(J_2=0.3,{\bf k}=\left(\pi,0\right)\right) 
&=&	0.37 - \frac1n 0.03,\\
\label{weightpipim}
{\cal W}_m \left(J_2=0.3,{\bf k}=\left(\frac\pi2,\frac\pi2\right)\right) 
&=&	0.38 - \frac1n 0.04.
\end{eqnarray}

Also shown in Fig.~\ref{spec_j1j203} is the singlon spectrum found using a plaquette expansion (PE) up to the 7-th order in the inter-plaquette interaction by the method proposed in our previous papers \cite{singlon,singlon2}. Although PE is more suitable for discussion of disordered phases with singlet ground states, \cite{singlon,singlon2} the singlon spectrum obtained by PE shows a good quantitative agreement with BOT results even in the ordered phase not far from transition points (as it is seen from Fig.~\ref{spec_j1j203} and as it will be shown in our forthcoming paper \cite{weunp} devoted to the disordered phase in the $J_1$--$J_2$ model).

It is also seen from Fig.~\ref{spec_j1j203} that other spin-0 and spin-1 excitations (denoted as "bound states" in Figs.~\ref{spec_ss} and \ref{spec_j1j203}) become closer to each other and to the magnon spectrum. We demonstrate in our forthcoming paper \cite{weunp}, that these spin-0 and spin-1 branches merge in the disordered phase forming a high-energy triplon branch (in addition to the triplon branch stemming from the magnon and the Higgs modes) which plays an important role in the disordered phase. Fig.~\ref{spec_j1j203} demonstrates that spectra of these spin-0 and spin-1 excitations are particularly close to spectra of the Higgs mode and magnons at ${\bf k}=(\pi/2,\pi/2)$, where their damping is minimal. One expects also that the damping of these excitations is overestimated near ${\bf k}=(\pi/2,\pi/2)$ in the considered first order in $1/n$ because bare spectra are used to calculate it in this order: $1/n$ corrections decrease their energies and bring them closer to the lower edge of the two-magnon continuum thus providing less phase space for the decay. Thus, we predict that at sufficiently large $J_2$ extra anomalies can appear in DSFs stemming from these excitations. In particular, the spin-1 excitation gives a peak in the transverse DSF. However its spectral weight
\begin{equation}
\label{weightpipispin1}
{\cal W}_{bs}^{\rm spin-1} \left(J_2=0.3,{\bf k}=\left(\frac\pi2,\frac\pi2\right)\right) 
=	0.0129 + \frac1n 0.0003
\end{equation}
is more than an order of magnitude smaller than the magnon spectral weight (cf.\ Eq.~\eqref{weightpipim}). In contrast, spectral weights of anomalies in the longitudinal DSF from the spin-0 and the Higgs quasiparticles
\begin{eqnarray}
\label{weightpipispin0}
{\cal W}_{bs}^{\rm spin-0} \left(J_2=0.3,{\bf k}=\left(\frac\pi2,\frac\pi2\right)\right) 
&=&	0.055 - \frac1n 0.008,\\
\label{weightpipiHiggs}
{\cal W}_{\rm Higgs} \left(J_2=0.3,{\bf k} = \left(\frac\pi2,\frac\pi2\right)\right) 
&=&	0.059 - \frac1n 0.013
\end{eqnarray}
are quite comparable with the magnon spectral weight \eqref{weightpipim}. Then, the Higgs and the spin-0 excitations can be visible around ${\bf k}=(\pi/2,\pi/2)$ even in an inelastic neutron scattering experiment not distinguishing the transverse and the longitudinal channels.

\section{Summary and conclusion}
\label{conc}

To summarize, using the four-spin BOT proposed in Ref.~\cite{ibot}, we discuss spectral properties of spin-$\frac12$ $J_1$--$J_2$ Heisenberg antiferromagnet \eqref{ham} on square lattice in the N\'eel phase (i.e., at $J_2<0.4$) and in the first order in $1/n$, where $n$ is the maximum number of bosons which can occupy a unit cell (physical results correspond to $n=1$). 

At $J_2=0$, the obtained magnon spectrum (see Fig.~\ref{spec_ss}) is in good quantitative agreement with experiment in CFTD even around ${\bf k}=(\pi,0)$. Calculated one-magnon spectral weights are in good quantitative agreement with the experiment except for the neighborhood of ${\bf k}=(\pi,0)$, where theoretical results overestimate the spectral weights (see Fig.~\ref{weights}). Besides, we do not observe the experimentally and numerically obtained pronounced high-energy continuum of excitations at ${\bf k}=(\pi,0)$ starting from the magnon peak. Thus, we do not support the idea suggested before \cite{pow1,pow2} that magnons with ${\bf k}=(\pi,0)$ are unstable with respect to the decay into another magnon and the Higgs excitation.

Upon $J_2$ increasing, one-magnon spectral weights decrease (see Fig.~\ref{weights03fig}) and the deviation around ${\bf k}=(\pi,0)$ becomes more pronounced of the magnon spectrum obtained using BOT from the spectrum observed in the second order in $1/S$ (see Fig.~\ref{spec_j1j203}). Spectra of all high-energy excitations move down and become closer to the magnon spectrum and to the lower edge of the two-magnon continuum. As a result, singlon (a spin-0 excitation responsible for the asymmetric peak in the Raman intensity in the $B_{1g}$ geometry) becomes a long-lived quasiparticle in the whole BZ and its spectrum merges with the magnon spectrum at $J_2\approx0.3$. Energies of the amplitude mode, another spin-0 excitation, and a spin-1 quasiparticle (which could appear in conventional approaches as a three-magnon bound state) become very close to the magnon energy at ${\bf k}=(\pi/2,\pi/2)$ and $J_2\agt0.3$ and their damping decreases. Then, these elementary excitations should produce distinct anomalies in the transverse and the longitudinal DSFs whose spectral weights are given by Eqs.~\eqref{weightpipispin1}--\eqref{weightpipiHiggs}. Experimental observation of the spin-1 excitation would be difficult, however, due to the smallness of its spectral weight in comparison with the magnon spectral weight given by Eq.~\eqref{weightpipim}.

\begin{acknowledgments}

This work is supported by Foundation for the advancement of theoretical physics and mathematics "BASIS".

\end{acknowledgments}

\appendix

\section{Dyson equations}
\label{app}

We present in this appendix sets of Dyson equations for Green's functions of spin-0 and spin-1 bosons. Determinants of these linear set of equations give denominators of the longitudinal and the transverse spin susceptibilities, correspondingly, discussed in the main text (Eq.~\eqref{denom}). We follow the notation introduced in Ref.~\cite{ibot}, where $a_{2,3,4,5}$, $b_{1,2,3,4}$, and $\tilde b_{1,2,3,4}$ are Bose operators describing excitations having projections on quantized axis 0, $+1$, and $-1$, respectively. Notice that we do not use the Bogoliubov transformation of operators. This approach is an extension of that used, e.g., in Ref.~\cite{syromyat}.

The set of Dyson equations for Green's functions of spin-0 elementary excitations has the form
\begin{eqnarray}
&&-G_{\left\{a_{2 \bf k},a^\dagger_{2 \bf k}\right\}} \left(\omega-S_{\left\{a^\dagger_{2 \bf k},a_{2 \bf k}\right\}}\right)
+G_{\left\{a^\dagger_{2 \bf -k},a^\dagger_{2 \bf k}\right\}} S_{\left\{a^\dagger_{2 \bf k},a^\dagger_{2 \bf -k}\right\}}+G_{\left\{a_{3 \bf k},a^\dagger_{2 \bf k}\right\}} S_{\left\{a^\dagger_{2 \bf k},a_{3 \bf k}\right\}}
+G_{\left\{a^\dagger_{3 \bf -k},a^\dagger_{2 \bf k}\right\}} S_{\left\{a^\dagger_{2 \bf k},a^\dagger_{3 \bf -k}\right\}}\nonumber\\
&&+G_{\left\{a_{4 \bf k},a^\dagger_{2 \bf k}\right\}} S_{\left\{a^\dagger_{2 \bf k},a_{4 \bf k}\right\}}+G_{\left\{a^\dagger_{4 \bf -k},a^\dagger_{2 \bf k}\right\}} S_{\left\{a^\dagger_{2 \bf k},a^\dagger_{4 \bf -k}\right\}}+G_{\left\{a_{5 \bf k},a^\dagger_{2 \bf k}\right\}} S_{\left\{a^\dagger_{2 \bf k},a_{5 \bf k}\right\}}+G_{\left\{a^\dagger_{5 \bf -k},a^\dagger_{2 \bf k}\right\}} S_{\left\{a^\dagger_{2 \bf k},a^\dagger_{5 \bf -k}\right\}}=-1,\nonumber\\
&&-G_{\left\{a_{3 \bf k},a^\dagger_{2 \bf k}\right\}} \left(\omega-S_{\left\{a^\dagger_{3 \bf k},a_{3 \bf k}\right\}}\right)+G_{\left\{a_{2 \bf k},a^\dagger_{2 \bf k}\right\}} S_{\left\{a^\dagger_{3 \bf k},a_{2 \bf k}\right\}}
+G_{\left\{a^\dagger_{2 \bf -k},a^\dagger_{2 \bf k}\right\}} S_{\left\{a^\dagger_{3 \bf k},a^\dagger_{2 \bf -k}\right\}}
+G_{\left\{a^\dagger_{3 \bf -k},a^\dagger_{2 \bf k}\right\}} S_{\left\{a^\dagger_{3 \bf k},a^\dagger_{3 \bf -k}\right\}}\nonumber\\
&&+G_{\left\{a_{4 \bf k},a^\dagger_{2 \bf k}\right\}} S_{\left\{a^\dagger_{3 \bf k},a_{4 \bf k}\right\}}+G_{\left\{a^\dagger_{4 \bf -k},a^\dagger_{2 \bf k}\right\}} S_{\left\{a^\dagger_{3 \bf k},a^\dagger_{4 \bf -k}\right\}}+G_{\left\{a_{5 \bf k},a^\dagger_{2 \bf k}\right\}} S_{\left\{a^\dagger_{3 \bf k},a_{5 \bf k}\right\}}+G_{\left\{a^\dagger_{5 \bf -k},a^\dagger_{2 \bf k}\right\}} S_{\left\{a^\dagger_{3 \bf k},a^\dagger_{5 \bf -k}\right\}}=0,\nonumber\\
&&-G_{\left\{a_{4 \bf k},a^\dagger_{2 \bf k}\right\}} \left(\omega-S_{\left\{a^\dagger_{4 \bf k},a_{4 \bf k}\right\}}\right)+G_{\left\{a_{2 \bf k},a^\dagger_{2 \bf k}\right\}} S_{\left\{a^\dagger_{4 \bf k},a_{2 \bf k}\right\}}
+G_{\left\{a^\dagger_{2 \bf -k},a^\dagger_{2 \bf k}\right\}} S_{\left\{a^\dagger_{4 \bf k},a^\dagger_{2 \bf -k}\right\}}
+G_{\left\{a_{3 \bf k},a^\dagger_{2 \bf k}\right\}} S_{\left\{a^\dagger_{4 \bf k},a_{3 \bf k}\right\}}\nonumber\\
&&+G_{\left\{a^\dagger_{3 \bf -k},a^\dagger_{2 \bf k}\right\}} S_{\left\{a^\dagger_{4 \bf k},a^\dagger_{3 \bf -k}\right\}}+G_{\left\{a^\dagger_{4 \bf -k},a^\dagger_{2 \bf k}\right\}} S_{\left\{a^\dagger_{4 \bf k},a^\dagger_{4 \bf -k}\right\}}+G_{\left\{a_{5 \bf k},a^\dagger_{2 \bf k}\right\}} S_{\left\{a^\dagger_{4 \bf k},a_{5 \bf k}\right\}}+G_{\left\{a^\dagger_{5 \bf -k},a^\dagger_{2 \bf k}\right\}} S_{\left\{a^\dagger_{4 \bf k},a^\dagger_{5 \bf -k}\right\}}=0,\nonumber\\
&&-G_{\left\{a_{5 \bf k},a^\dagger_{2 \bf k}\right\}} \left(\omega-S_{\left\{a^\dagger_{5 \bf k},a_{5 \bf k}\right\}}\right)+G_{\left\{a_{2 \bf k},a^\dagger_{2 \bf k}\right\}} S_{\left\{a^\dagger_{5 \bf k},a_{2 \bf k}\right\}}
+G_{\left\{a^\dagger_{2 \bf -k},a^\dagger_{2 \bf k}\right\}} S_{\left\{a^\dagger_{5 \bf k},a^\dagger_{2 \bf -k}\right\}}
+G_{\left\{a_{3 \bf k},a^\dagger_{2 \bf k}\right\}} S_{\left\{a^\dagger_{5 \bf k},a_{3 \bf k}\right\}}\nonumber\\
&&+G_{\left\{a^\dagger_{3 \bf -k},a^\dagger_{2 \bf k}\right\}} S_{\left\{a^\dagger_{5 \bf k},a^\dagger_{3 \bf -k}\right\}}+G_{\left\{a_{4 \bf k},a^\dagger_{2 \bf k}\right\}} S_{\left\{a^\dagger_{5 \bf k},a_{4 \bf k}\right\}}+G_{\left\{a^\dagger_{4 \bf -k},a^\dagger_{2 \bf k}\right\}} S_{\left\{a^\dagger_{5 \bf k},a^\dagger_{4 \bf -k}\right\}}+G_{\left\{a^\dagger_{5 \bf -k},a^\dagger_{2 \bf k}\right\}} S_{\left\{a^\dagger_{5 \bf k},a^\dagger_{5 \bf -k}\right\}}=0,\\
&&G_{\left\{a^\dagger_{2 \bf -k},a^\dagger_{2 \bf k}\right\}} \left(\omega+S_{\left\{a_{2 \bf -k},a^\dagger_{2 \bf -k}\right\}}\right)+G_{\left\{a_{2 \bf k},a^\dagger_{2 \bf k}\right\}} S_{\left\{a_{2 \bf -k},a_{2 \bf k}\right\}}
+G_{\left\{a_{3 \bf k},a^\dagger_{2 \bf k}\right\}} S_{\left\{a_{2 \bf -k},a_{3 \bf k}\right\}}
+G_{\left\{a^\dagger_{3 \bf -k},a^\dagger_{2 \bf k}\right\}} S_{\left\{a_{2 \bf -k},a^\dagger_{3 \bf -k}\right\}}\nonumber\\
&&+G_{\left\{a_{4 \bf k},a^\dagger_{2 \bf k}\right\}} S_{\left\{a_{2 \bf -k},a_{4 \bf k}\right\}}+G_{\left\{a^\dagger_{4 \bf -k},a^\dagger_{2 \bf k}\right\}} S_{\left\{a_{2 \bf -k},a^\dagger_{4 \bf -k}\right\}}+G_{\left\{a_{5 \bf k},a^\dagger_{2 \bf k}\right\}} S_{\left\{a_{2 \bf -k},a_{5 \bf k}\right\}}+G_{\left\{a^\dagger_{5 \bf -k},a^\dagger_{2 \bf k}\right\}} S_{\left\{a_{2 \bf -k},a^\dagger_{5 \bf -k}\right\}}=0,\nonumber\\
&&G_{\left\{a^\dagger_{3 \bf -k},a^\dagger_{2 \bf k}\right\}} \left(\omega+S_{\left\{a_{3 \bf -k},a^\dagger_{3 \bf -k}\right\}}\right)+G_{\left\{a_{2 \bf k},a^\dagger_{2 \bf k}\right\}} S_{\left\{a_{3 \bf -k},a_{2 \bf k}\right\}}
+G_{\left\{a^\dagger_{2 \bf -k},a^\dagger_{2 \bf k}\right\}} S_{\left\{a_{3 \bf -k},a^\dagger_{2 \bf -k}\right\}}
+G_{\left\{a_{3 \bf k},a^\dagger_{2 \bf k}\right\}} S_{\left\{a_{3 \bf -k},a_{3 \bf k}\right\}}\nonumber\\
&&+G_{\left\{a_{4 \bf k},a^\dagger_{2 \bf k}\right\}} S_{\left\{a_{3 \bf -k},a_{4 \bf k}\right\}}+G_{\left\{a^\dagger_{4 \bf -k},a^\dagger_{2 \bf k}\right\}} S_{\left\{a_{3 \bf -k},a^\dagger_{4 \bf -k}\right\}}+G_{\left\{a_{5 \bf k},a^\dagger_{2 \bf k}\right\}} S_{\left\{a_{3 \bf -k},a_{5 \bf k}\right\}}+G_{\left\{a^\dagger_{5 \bf -k},a^\dagger_{2 \bf k}\right\}} S_{\left\{a_{3 \bf -k},a^\dagger_{5 \bf -k}\right\}}=0,\nonumber\\
&&G_{\left\{a^\dagger_{4 \bf -k},a^\dagger_{2 \bf k}\right\}} \left(\omega+S_{\left\{a_{4 \bf -k},a^\dagger_{4 \bf -k}\right\}}\right)+G_{\left\{a_{2 \bf k},a^\dagger_{2 \bf k}\right\}} S_{\left\{a_{4 \bf -k},a_{2 \bf k}\right\}}
+G_{\left\{a^\dagger_{2 \bf -k},a^\dagger_{2 \bf k}\right\}} S_{\left\{a_{4 \bf -k},a^\dagger_{2 \bf -k}\right\}}
+G_{\left\{a_{3 \bf k},a^\dagger_{2 \bf k}\right\}} S_{\left\{a_{4 \bf -k},a_{3 \bf k}\right\}}\nonumber\\
&&+G_{\left\{a^\dagger_{3 \bf -k},a^\dagger_{2 \bf k}\right\}} S_{\left\{a_{4 \bf -k},a^\dagger_{3 \bf -k}\right\}}+G_{\left\{a_{4 \bf k},a^\dagger_{2 \bf k}\right\}} S_{\left\{a_{4 \bf -k},a_{4 \bf k}\right\}}+G_{\left\{a_{5 \bf k},a^\dagger_{2 \bf k}\right\}} S_{\left\{a_{4 \bf -k},a_{5 \bf k}\right\}}+G_{\left\{a^\dagger_{5 \bf -k},a^\dagger_{2 \bf k}\right\}} S_{\left\{a_{4 \bf -k},a^\dagger_{5 \bf -k}\right\}}=0,\nonumber\\
&&G_{\left\{a^\dagger_{5 \bf -k},a^\dagger_{2 \bf k}\right\}} \left(\omega+S_{\left\{a_{5 \bf -k},a^\dagger_{5 \bf -k}\right\}}\right)+G_{\left\{a_{2 \bf k},a^\dagger_{2 \bf k}\right\}} S_{\left\{a_{5 \bf -k},a_{2 \bf k}\right\}}
+G_{\left\{a^\dagger_{2 \bf -k},a^\dagger_{2 \bf k}\right\}} S_{\left\{a_{5 \bf -k},a^\dagger_{2 \bf -k}\right\}}
+G_{\left\{a_{3 \bf k},a^\dagger_{2 \bf k}\right\}} S_{\left\{a_{5 \bf -k},a_{3 \bf k}\right\}}\nonumber\\
&&+G_{\left\{a^\dagger_{3 \bf -k},a^\dagger_{2 \bf k}\right\}} S_{\left\{a_{5 \bf -k},a^\dagger_{3 \bf -k}\right\}}+G_{\left\{a_{4 \bf k},a^\dagger_{2 \bf k}\right\}} S_{\left\{a_{5 \bf -k},a_{4 \bf k}\right\}}+G_{\left\{a^\dagger_{4 \bf -k},a^\dagger_{2 \bf k}\right\}} S_{\left\{a_{5 \bf -k},a^\dagger_{4 \bf -k}\right\}}+G_{\left\{a_{5 \bf k},a^\dagger_{2 \bf k}\right\}} S_{\left\{a_{5 \bf -k},a_{5 \bf k}\right\}}=0,\nonumber
\end{eqnarray}
where $G_{\left\{A,B\right\}}$ is the Green's function of operators $A$ and $B$, $S_{\left\{A,B\right\}}=C_{AB}+\Sigma_{\left\{A,B\right\}}(\omega,{\bf k})$, $\Sigma_{\left\{A,B\right\}}(\omega,{\bf k})$ is the self-energy part, and $C_{AB}$ is the coefficient of the term in the bilinear part of the Hamiltonian ${\cal H}_2$ containing the product $AB$. 
We do not present numerous coefficients $C_{AB}$ here (${\cal H}_2$  contains 103 terms at $J_2\ne0$). We will provide them on request.

Dyson equations for Green's functions of spin-1 excitations have the form
\begin{eqnarray}
&&-G_{\left\{{b}_{1\bf k},b^\dagger_{1\bf k}\right\}} \left(\omega-S_{\left\{b^\dagger_{1\bf k},{b}_{1\bf k}\right\}}\right)
+G_{\left\{{b}_{2\bf k},b^\dagger_{1\bf k}\right\}} S_{\left\{b^\dagger_{1\bf k},{b}_{2\bf k}\right\}}
+G_{\left\{{b}_{3\bf k},b^\dagger_{1\bf k}\right\}} S_{\left\{b^\dagger_{1\bf k},{b}_{3\bf k}\right\}}
+G_{\left\{{b}_{4\bf k},b^\dagger_{1\bf k}\right\}} S_{\left\{b^\dagger_{1\bf k},{b}_{4\bf k}\right\}}\nonumber\\
&&+G_{\left\{\tilde b^\dagger_{1\bf -k},b^\dagger_{1\bf k}\right\}} S_{\left\{b^\dagger_{1\bf k},\tilde b^\dagger_{1\bf -k}\right\}}
+G_{\left\{\tilde b^\dagger_{2\bf -k},b^\dagger_{1\bf k}\right\}} S_{\left\{b^\dagger_{1\bf k},\tilde b^\dagger_{2\bf -k}\right\}}
+G_{\left\{\tilde b^\dagger_{3\bf -k},b^\dagger_{1\bf k}\right\}} S_{\left\{b^\dagger_{1\bf k},\tilde b^\dagger_{3\bf -k}\right\}}
+G_{\left\{\tilde b^\dagger_{4\bf -k},b^\dagger_{1\bf k}\right\}} S_{\left\{b^\dagger_{1\bf k},\tilde b^\dagger_{4\bf -k}\right\}}=-1,\nonumber\\
&&-G_{\left\{{b}_{2\bf k},b^\dagger_{1\bf k}\right\}} \left(\omega-S_{\left\{b^\dagger_{2\bf k},{b}_{2\bf k}\right\}}\right)
+G_{\left\{{b}_{1\bf k},b^\dagger_{1\bf k}\right\}} S_{\left\{b^\dagger_{2\bf k},{b}_{1\bf k}\right\}}
+G_{\left\{{b}_{3\bf k},b^\dagger_{1\bf k}\right\}} S_{\left\{b^\dagger_{2\bf k},{b}_{3\bf k}\right\}}
+G_{\left\{{b}_{4\bf k},b^\dagger_{1\bf k}\right\}} S_{\left\{b^\dagger_{2\bf k},{b}_{4\bf k}\right\}}\nonumber\\
&&
+G_{\left\{\tilde b^\dagger_{1\bf -k},b^\dagger_{1\bf k}\right\}} S_{\left\{b^\dagger_{2\bf k},\tilde b^\dagger_{1\bf -k}\right\}}
+G_{\left\{\tilde b^\dagger_{2\bf -k},b^\dagger_{1\bf k}\right\}} S_{\left\{b^\dagger_{2\bf k},\tilde b^\dagger_{2\bf -k}\right\}}
+G_{\left\{\tilde b^\dagger_{3\bf -k},b^\dagger_{1\bf k}\right\}} S_{\left\{b^\dagger_{2\bf k},\tilde b^\dagger_{3\bf -k}\right\}}
+G_{\left\{\tilde b^\dagger_{4\bf -k},b^\dagger_{1\bf k}\right\}} S_{\left\{b^\dagger_{2\bf k},\tilde b^\dagger_{4\bf -k}\right\}}=0,\nonumber\\
&&-G_{\left\{{b}_{3\bf k},b^\dagger_{1\bf k}\right\}} \left(\omega-S_{\left\{b^\dagger_{3\bf k},{b}_{3\bf k}\right\}}\right)
+G_{\left\{{b}_{1\bf k},b^\dagger_{1\bf k}\right\}} S_{\left\{b^\dagger_{3\bf k},{b}_{1\bf k}\right\}}
+G_{\left\{{b}_{2\bf k},b^\dagger_{1\bf k}\right\}} S_{\left\{b^\dagger_{3\bf k},{b}_{2\bf k}\right\}}
+G_{\left\{{b}_{4\bf k},b^\dagger_{1\bf k}\right\}} S_{\left\{b^\dagger_{3\bf k},{b}_{4\bf k}\right\}}\nonumber\\
&&
+G_{\left\{\tilde b^\dagger_{1\bf -k},b^\dagger_{1\bf k}\right\}} S_{\left\{b^\dagger_{3\bf k},\tilde b^\dagger_{1\bf -k}\right\}}
+G_{\left\{\tilde b^\dagger_{2\bf -k},b^\dagger_{1\bf k}\right\}} S_{\left\{b^\dagger_{3\bf k},\tilde b^\dagger_{2\bf -k}\right\}}
+G_{\left\{\tilde b^\dagger_{3\bf -k},b^\dagger_{1\bf k}\right\}} S_{\left\{b^\dagger_{3\bf k},\tilde b^\dagger_{3\bf -k}\right\}}
+G_{\left\{\tilde b^\dagger_{4\bf -k},b^\dagger_{1\bf k}\right\}} S_{\left\{b^\dagger_{3\bf k},\tilde b^\dagger_{4\bf -k}\right\}}=0,\\
&&-G_{\left\{{b}_{4\bf k},b^\dagger_{1\bf k}\right\}} \left(\omega-S_{\left\{b^\dagger_{4\bf k},{b}_{4\bf k}\right\}}\right)
+G_{\left\{{b}_{1\bf k},b^\dagger_{1\bf k}\right\}} S_{\left\{b^\dagger_{4\bf k},{b}_{1\bf k}\right\}}
+G_{\left\{{b}_{2\bf k},b^\dagger_{1\bf k}\right\}} S_{\left\{b^\dagger_{4\bf k},{b}_{2\bf k}\right\}}
+G_{\left\{{b}_{3\bf k},b^\dagger_{1\bf k}\right\}} S_{\left\{b^\dagger_{4\bf k},{b}_{3\bf k}\right\}}\nonumber\\
&&
+G_{\left\{\tilde b^\dagger_{1\bf -k},b^\dagger_{1\bf k}\right\}} S_{\left\{b^\dagger_{4\bf k},\tilde b^\dagger_{1\bf -k}\right\}}
+G_{\left\{\tilde b^\dagger_{2\bf -k},b^\dagger_{1\bf k}\right\}} S_{\left\{b^\dagger_{4\bf k},\tilde b^\dagger_{2\bf -k}\right\}}
+G_{\left\{\tilde b^\dagger_{3\bf -k},b^\dagger_{1\bf k}\right\}} S_{\left\{b^\dagger_{4\bf k},\tilde b^\dagger_{3\bf -k}\right\}}
+G_{\left\{\tilde b^\dagger_{4\bf -k},b^\dagger_{1\bf k}\right\}} S_{\left\{b^\dagger_{4\bf k},\tilde b^\dagger_{4\bf -k}\right\}}=0,\nonumber\\
&&G_{\left\{\tilde b^\dagger_{1\bf -k},b^\dagger_{1\bf k}\right\}} \left(\omega+S_{\left\{\tilde b_{1\bf -k},\tilde b^\dagger_{1\bf -k}\right\}}\right)
+G_{\left\{{b}_{1\bf k},b^\dagger_{1\bf k}\right\}} S_{\left\{\tilde b_{1\bf -k},{b}_{1\bf k}\right\}}
+G_{\left\{{b}_{2\bf k},b^\dagger_{1\bf k}\right\}} S_{\left\{\tilde b_{1\bf -k},{b}_{2\bf k}\right\}}
+G_{\left\{{b}_{3\bf k},b^\dagger_{1\bf k}\right\}} S_{\left\{\tilde b_{1\bf -k},{b}_{3\bf k}\right\}}\nonumber\\
&&
+G_{\left\{{b}_{4\bf k},b^\dagger_{1\bf k}\right\}} S_{\left\{\tilde b_{1\bf -k},{b}_{4\bf k}\right\}}
+G_{\left\{\tilde b^\dagger_{2\bf -k},b^\dagger_{1\bf k}\right\}} S_{\left\{\tilde b_{1\bf -k},\tilde b^\dagger_{2\bf -k}\right\}}
+G_{\left\{\tilde b^\dagger_{3\bf -k},b^\dagger_{1\bf k}\right\}} S_{\left\{\tilde b_{1\bf -k},\tilde b^\dagger_{3\bf -k}\right\}}
+G_{\left\{\tilde b^\dagger_{4\bf -k},b^\dagger_{1\bf k}\right\}} S_{\left\{\tilde b_{1\bf -k},\tilde b^\dagger_{4\bf -k}\right\}}=0,\nonumber\\
&&G_{\left\{\tilde b^\dagger_{2\bf -k},b^\dagger_{1\bf k}\right\}} \left(\omega+S_{\left\{\tilde b_{2\bf -k},\tilde b^\dagger_{2\bf -k}\right\}}\right)
+G_{\left\{{b}_{1\bf k},b^\dagger_{1\bf k}\right\}} S_{\left\{\tilde b_{2\bf -k},{b}_{1\bf k}\right\}}
+G_{\left\{{b}_{2\bf k},b^\dagger_{1\bf k}\right\}} S_{\left\{\tilde b_{2\bf -k},{b}_{2\bf k}\right\}}
+G_{\left\{{b}_{3\bf k},b^\dagger_{1\bf k}\right\}} S_{\left\{\tilde b_{2\bf -k},{b}_{3\bf k}\right\}}\nonumber\\
&&
+G_{\left\{{b}_{4\bf k},b^\dagger_{1\bf k}\right\}} S_{\left\{\tilde b_{2\bf -k},{b}_{4\bf k}\right\}}
+G_{\left\{\tilde b^\dagger_{1\bf -k},b^\dagger_{1\bf k}\right\}} S_{\left\{\tilde b_{2\bf -k},\tilde b^\dagger_{1\bf -k}\right\}}
+G_{\left\{\tilde b^\dagger_{3\bf -k},b^\dagger_{1\bf k}\right\}} S_{\left\{\tilde b_{2\bf -k},\tilde b^\dagger_{3\bf -k}\right\}}
+G_{\left\{\tilde b^\dagger_{4\bf -k},b^\dagger_{1\bf k}\right\}} S_{\left\{\tilde b_{2\bf -k},\tilde b^\dagger_{4\bf -k}\right\}}=0,\nonumber\\
&&G_{\left\{\tilde b^\dagger_{3\bf -k},b^\dagger_{1\bf k}\right\}} \left(\omega+S_{\left\{\tilde b_{3\bf -k},\tilde b^\dagger_{3\bf -k}\right\}}\right)
+G_{\left\{{b}_{1\bf k},b^\dagger_{1\bf k}\right\}} S_{\left\{\tilde b_{3\bf -k},{b}_{1\bf k}\right\}}
+G_{\left\{{b}_{2\bf k},b^\dagger_{1\bf k}\right\}} S_{\left\{\tilde b_{3\bf -k},{b}_{2\bf k}\right\}}
+G_{\left\{{b}_{3\bf k},b^\dagger_{1\bf k}\right\}} S_{\left\{\tilde b_{3\bf -k},{b}_{3\bf k}\right\}}\nonumber\\
&&
+G_{\left\{{b}_{4\bf k},b^\dagger_{1\bf k}\right\}} S_{\left\{\tilde b_{3\bf -k},{b}_{4\bf k}\right\}}
+G_{\left\{\tilde b^\dagger_{1\bf -k},b^\dagger_{1\bf k}\right\}} S_{\left\{\tilde b_{3\bf -k},\tilde b^\dagger_{1\bf -k}\right\}}
+G_{\left\{\tilde b^\dagger_{2\bf -k},b^\dagger_{1\bf k}\right\}} S_{\left\{\tilde b_{3\bf -k},\tilde b^\dagger_{2\bf -k}\right\}}
+G_{\left\{\tilde b^\dagger_{4\bf -k},b^\dagger_{1\bf k}\right\}} S_{\left\{\tilde b_{3\bf -k},\tilde b^\dagger_{4\bf -k}\right\}}=0,\nonumber\\
&&G_{\left\{\tilde b^\dagger_{4\bf -k},b^\dagger_{1\bf k}\right\}} \left(\omega+S_{\left\{\tilde b_{4\bf -k},\tilde b^\dagger_{4\bf -k}\right\}}\right)
+G_{\left\{{b}_{1\bf k},b^\dagger_{1\bf k}\right\}} S_{\left\{\tilde b_{4\bf -k},{b}_{1\bf k}\right\}}
+G_{\left\{{b}_{2\bf k},b^\dagger_{1\bf k}\right\}} S_{\left\{\tilde b_{4\bf -k},{b}_{2\bf k}\right\}}
+G_{\left\{{b}_{3\bf k},b^\dagger_{1\bf k}\right\}} S_{\left\{\tilde b_{4\bf -k},{b}_{3\bf k}\right\}}\nonumber\\
&&
+G_{\left\{{b}_{4\bf k},b^\dagger_{1\bf k}\right\}} S_{\left\{\tilde b_{4\bf -k},{b}_{4\bf k}\right\}}
+G_{\left\{\tilde b^\dagger_{1\bf -k},b^\dagger_{1\bf k}\right\}} S_{\left\{\tilde b_{4\bf -k},\tilde b^\dagger_{1\bf -k}\right\}}
+G_{\left\{\tilde b^\dagger_{2\bf -k},b^\dagger_{1\bf k}\right\}} S_{\left\{\tilde b_{4\bf -k},\tilde b^\dagger_{2\bf -k}\right\}}
+G_{\left\{\tilde b^\dagger_{3\bf -k},b^\dagger_{1\bf k}\right\}} S_{\left\{\tilde b_{4\bf -k},\tilde b^\dagger_{3\bf -k}\right\}}=0.	\nonumber
\end{eqnarray}

\bibliography{j1j2bib}

\end{document}